\documentclass[twocolumn,showpacs,preprintnumbers,amsmath,amssymb]{revtex4}

\usepackage{graphicx}
\usepackage{dcolumn}
\usepackage{bm}

\begin{document}

\title{Fluidization of wet granular matter}

\author{Dimitrios Geromichalos}
    \email{Dimitrios.Geromichalos@physik.uni-ulm.de}

\author{Mario Scheel}

\author{Stephan Herminghaus}

\affiliation{%
    Applied Physics Lab., Ulm University, Ulm, D-89069,
    Germany
}

\date{\today}

\begin{abstract}
We have studied the effect of small amounts of added liquid on the
dynamic behavior of a granular system consisting of spherical
glass beads. The critical acceleration $\Gamma$ for fluidization
is found to increase strongly when liquid is added. It furthermore
depends on the bead radius, the height of the sample, and the
shaking frequency. As our main result, we find that these
dependencies can be presented in a factorized form. We present a
simple model which accounts for most of our findings.
\end{abstract}


\maketitle

As it is generally known from everyday experience, the mechanical
properties of a granular material change dramatically if some
liquid is added. The main reason is the internal cohesion due to
capillary forces arising from liquid bridges between the grains
\cite{mik98, israelachvili, horn97, hal98, boc98, fra99}. While
recent years have seen considerable progress in understanding the
dynamics of dry granular materials \cite{ristow, jae96, kad99,
dix03}, the physical mechanisms underlying the properties of wet
systems still pose many unanswered questions.

In this article we cosider granular matter in a container vibrated
vertically with a certain frequency $\nu$ and an amplitude $A$
\cite{ristow}. If this agitation is weak, the particles of the
granulate remain fixed and do not move against each other. When
the strength of the agitation is increased beyond the so-called
{\it fluidization threshold} by enhancing $A$ or $\nu$, the
granulate becomes {\it fluidized} \cite{ristow}. In this state the
cohesiveness of the granulate is overcome. This enables a
liquid-like flow of the granular matter, e.g. through pipe systems
\cite{duran}. Furthermore there starts in general a convective
flow resembling to that of a fluid. The form of the vibration used
in most experiments was a sinusoidal horizontal or, in particular,
a vertical vibration. \cite{ristow, ris97, met02}

In our experiments, we have studied the effect of capillary
forces, which are characteristic for wet granular matter, on the
fluidization behavior of the granulate \footnote{The only
fluidization experiments which are mentioned in literature and
concern wet granular matter were done by blowing air through the
granulate \cite{sim93} and are therefore not immediately
comparable to the experiments presented in this paper.}. We
mounted a small cylindrical glass container (diameter: $2.5$ $cm$;
height: $4$ $cm$) filled at about three quarters with the granular
matter consisting of glass beads on an inductive shaker. The
container was shaken vertically with amplitudes between $1 \mu m$
and $1 mm$ and frequencies between $20$ and $333 Hz$. We conducted
the experiment with several samples composed of particles with
radii from $138 \mu m$ to $500 \mu m$, respectively. In order to
prevent crystallization, the beads in each sample were chosen to
be slightly polydisperse: the spread in bead size ranged from 10
to 20 $\%$. We added controlled amounts of liquid and shook the
sample by hand for some minutes in order to obtain a homogeneous
distribution of liquid.

For the results presented here, we used water as the liquid, but
similar results were obtained using nonane. We can therefore
exclude that leaching of ions from the glass, and other side
effects due to the specific properties of water \cite{oli02}, play
a significant role.

We measured the fluidization threshold as follows: The shaking
frequency $\nu$ was fixed at a certain value. The shaking
amplitude $A$ was increased until a relative motion of the
particles was visible through the glass wall. It turned out that
this transition was rather well defined, since small variations of
less than $5 \%$ of the amplitude at the critical value were
decisive for the existence of the fluidization behavior. From the
amplitude and the frequency we calculated the fluidization
amplitude:
\[
\Gamma := 4 \pi^2 \frac{A \nu^2}{g}
\]
where $g$ is the gravitational acceleration.

We found that the fluidization acceleration for wet granular
matter depends strongly on the water content $W$, but also on the
bead radius $R$, the sample height $H$ and the shaking frequency
$\nu$ as follows:
\[
\Gamma (W, H, R, \nu) = \Gamma_0 (1 + G(W, H, R, \nu))
\]
with $\Gamma_0 = 1.2$ \cite{ristow} being the value for the dry
case. As it turned out, the function $G(W, H, R, \nu)$ can be
factorized and written as
\begin{equation}
\label{factorize:eq} G(W, H, R, \nu) = f (W) h (H) p (R) q (\nu)
\end{equation}
i.e. the contribution of each single parameter can be treated
separately. This is the main result of this paper. The form of the
dependencies is described below.

\begin{figure}[h]
 \includegraphics[width = 8.5cm]{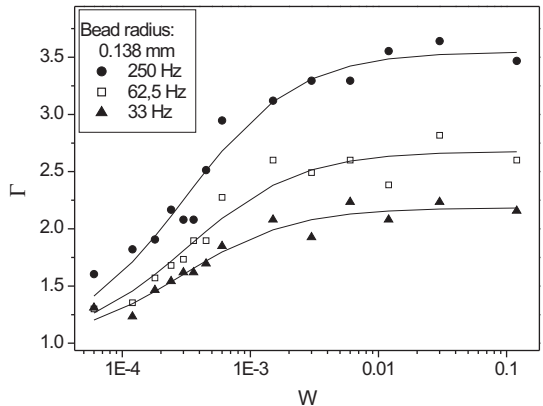}
 \caption{\label{gamma-w:pict} Dependence of the fluidization acceleration for
 different frequencies for $R = 0.138 mm$ on the water content.
 The fitting functions are described in the text.}
\end{figure}

{\it Dependence on water content:} The fluidization amplitude
$\Gamma$ was found to depend strongly on the water content $W$ as
shown in Fig. \ref{gamma-w:pict}. For small water contents the
fluidization acceleration $\Gamma$ was a rapidly increasing but
smooth function of $W$ and went from the value for the dry
granulate ($W=0$) to a plateau value. The value of the water
content at which this plateau was reached, $W_b$, was roughly $0.1
\%$, as is shown in Fig. \ref{gamma-w:pict} for different
frequencies. $\Gamma$ saturated then at this plateau value and
stayed constant till water contents of the order of magnitude of
$10 \%$. The plateau value depended on the frequency $\nu$ and the
bead radius $R$, as we will discuss later in the text.

We can provide an explanation for this behavior in terms of the
capillary forces which have a hysteretic character \cite{wil00,
cra94, lothar02, schulz03}.

For this we look at first at the capillary forces between two
glass beads with a radius $R$. As it can be shown analytically
\cite{wil00}, the capillary force between two perfectly smooth
beads depends for sufficient small liquid contents only on the
radius $R$, but not on $W$:
\[
F_{cap} = 2 \pi R \gamma
\]
with the surface tension $\gamma$.

On the other hand it was found experimentally \cite{rumpf62} that
for very small water contents and beads with a rough surface, the
capillary forces are not constant but decrease monotonically and
smoothly to zero for $W \rightarrow 0$:
\[
F_{cap} = 2 \pi R \gamma f(w)
\]
with the volume of a single capillary bridge $w$ and $f$ being a
function which depends on the roughness of the sphere surfaces
\cite{hal98, wil00}.

The function $f(w)$ is for perfect spheres and complete wetting
equal to one but for a rough surface, it tends to zero as $w
\rightarrow 0$. The shape of $f(w)$ has been discussed in detail
before \cite{hal98}. Based on these considerations, it may be
approximated by
\begin{equation}
f(w)=\left[\frac{w}{w+w_0}\right]^{\mu}
\end{equation}
with $w_0 := R \delta ^2$ characterizing the roughness amplitude,
$\delta$. For the exponent, we have $\mu = (2-\chi)/(2+\chi)$,
where $\chi$ is the roughness exponent of the bead surface. For
(typical) scratch and dig roughness as for our beads, $\chi$ is
close to zero, and $\mu \approx 1$. This corresponds to the regime
found by Hornbaker et al. \cite{horn97}.

At equilibrium, we assume that all of the liquid is in the
bridges, $w = \frac{8 \pi R^3 W}{3 c \rho_{p} }$, where $c$ is the
coordination number of the network of bridges ($c \approx 6$)
\cite{ger03a} and $\rho_p$ is the packing density of the spheres
$\rho_p \approx 0.64$. Under dynamic conditions the typical bridge
volume would be expected to be somewhat smaller than the
equilibrium value. Our own measurements \cite{koh03} suggest that
$w = \alpha R^3 W$, where $\alpha \approx 0.25$ \cite{ger03}.

We assume that the starting condition for fluidization due to {\it
vertical} shaking is that the force $F_0$ exerted from the
vibration is big enough to break the capillary bridges in at least
one {\it horizontal} plane in the pile. Here the vibrational force
is given directly by Newton's law:
\begin{equation}
\label{newton-force:eq} F_0 = (\Gamma - \Gamma_0) M g
\end{equation}
with the mass of the granular sample $M$ and $\Gamma_0 := \Gamma
(W=0) \approx 1.2$ \cite{ristow}. The mass of the (cylindrical)
pile is given by
\[
M = \rho_g \rho_{rcp} H B
\]
with the density of the bead material (glass) $\rho_g$, the random
close packing density for beads $\rho_{rcp} \approx 0.64$
\cite{sco60}, the height of the cylindrical sample $H$ and its
base area $B$.

The expression for the force exerted from the capillaries in the
relevant horizontal plane is
\begin{equation}
\label{plane-force:eq} F_{pl} = \frac{B}{R^2} \rho_{2D} 2 R \gamma
f(w)
\end{equation}
with the two-dimendional 'density' $\rho_{2D}$ of the bridges
which have to break.

From equating (\ref{newton-force:eq}) and (\ref{plane-force:eq}),
one gets

\[
\Gamma (W) = \Gamma_0 \left( 1 +  \frac{\widetilde{c} f(w)}{H R}
\right)
\]
with $\widetilde{c} = \frac{2 \gamma \rho_{2D}}{\Gamma_0 \rho_g
\rho_{rcp} g}$. As it can be seen in Fig. \ref{gamma-w:pict}, this
can be used quite well to describe the $W$-dependence of $\Gamma$.
$\widetilde{c}$ was a fitting parameter which was also a function
of the bead radius $R$ and the shaking frequency $\nu$ (see
below).

The function $f(w)$ connects the bridge formation directly to the
roughness. From $W_0 \approx 0.05 \%$ (see Fig.
\ref{gamma-w:pict}) we get $\delta \sim 500nm$, which is similar
to the peak-to-peak roughness obtained from the inspection of the
beads by atomic force microscopy \footnote{Furthermore we should
add that the capillary forces depend strongly on the particle
shape as we saw when we did similar experiments with NaCl-cubes
and nonane as wetting liquid and found no saturation.}. As it can
be seen in Fig. \ref{gamma-w:pict}, the dependence of the
fluidization acceleration on the liquid content is described quite
well by this theory.

\begin{figure}[h]
 \includegraphics[width = 8.5cm]{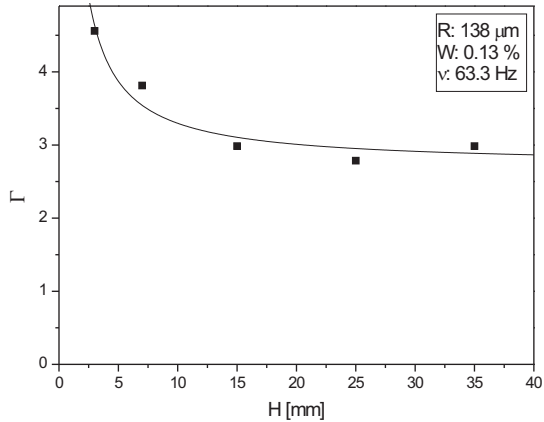}
 \caption{\label{gamma-h:pict} Dependence of the fluidization acceleration on the height $H$ of the granular pile.
 The fitting function is described in the text ($H_c \approx 4 mm$).}
\end{figure}

A further prediction is the dependence of $\Gamma$ on $H$, which
we could confirm at least qualitatively in additional control
measurements at $0.25 mm \leq H \leq 35 mm$ with $R = 138 \mu m$
and $W = 0.13 \%$ as is shown in Fig. \ref{gamma-h:pict}. We could
could fit the values plotted in Fig. \ref{gamma-h:pict} quite well
with
\begin{equation}
\label{gamma-wr-2:eq} \Gamma (W, H) = \Gamma_0 \left( 1 +
\frac{\widetilde{c} f(w)}{R} \left( \frac{1}{H} + \frac{1}{H_c}
\right) \right)
\end{equation}
The finite limiting value $H_c$ may be understood as follows. One
ruptured horizontal plane may be sufficient to fluidize the
surrounding medium only to a certain vertical distance. Hence in
samples with large $H$, the initial fluidization occurs in
horizontal layers with a thickness $H_c < H$ such that $\Gamma$
becomes independent of $H$. We were able observe this behavior by
direct optical monitoring.

\begin{figure}[h]
 \includegraphics[width = 8.5cm]{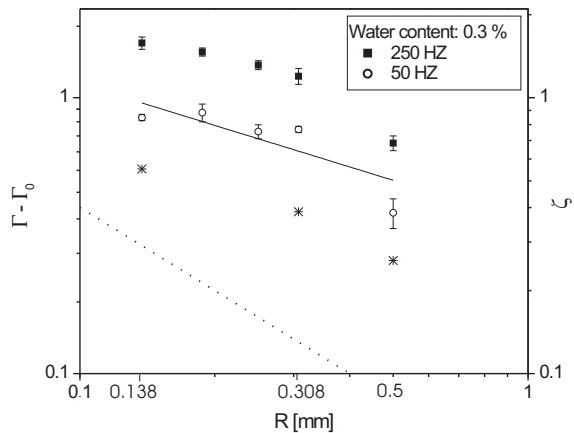}
 \caption{\label{gamma-r:pict}
 Bead radius dependence of $\Delta \Gamma := \Gamma (W, R) - \Gamma_0$ for
 different frequencies (circles and squares). \newline
 The asterisks represent the radius dependence of the slope $\zeta$ (described in the
 text; see Fig. \ref{gamma-nu:pict}). \newline
 The solid line is proportional to $R^{-\beta}$ with $\beta =
 0.5$,
 the dotted line is proportional to $R^{-1}$.}
\end{figure}

{\it Dependence on the bead radius:} As already mentioned the
dependence of $\Gamma$ on the bead radius $R$ proved to be
non-trivial. We found $\Gamma (W,R) - \Gamma_0$ to be proportional
to $\frac{1}{R^\beta}$ with $\beta \approx 0.5$, as shown in Fig.
\ref{gamma-r:pict} for $H \approx 3 cm$. We were able to describe
the data with
\begin{equation}
\label{gamma-wr:eq} \Gamma (W, R) = \Gamma_0 \left( 1 +
\frac{\widetilde{c} f(w)}{R^{\beta} R_0^{1 - \beta}} \left(
\frac{1}{H} + \frac{1}{H_c} \right) \right)
\end{equation}
The deviation of the measured values from the prediction according
to Eq. \ref{gamma-wr-2:eq} gets larger for smaller and smaller
bead radii. Hence we set $R_0 = 0.5 mm$ as the value where
\ref{gamma-wr-2:eq} and \ref{gamma-wr:eq} yield the same value for
$\Gamma$ \footnote{For $R < R_0$ the measured fluidization
acceleration according is smaller than the value obtained from
\ref{gamma-wr-2:eq}, since $\beta = 0.5 < 1$. Assuming that
particles with $R \geq R_0$ do not form clusters (explained in the
text), one would expect $\beta = 1$ for that case.}. Therewith we
obtained a $\rho_{2D}$ ranging between $3$ and $4$, which is a
reasonable result.

A possible explanation for $\beta \neq 1$ is that the particles of
a wet granulate tend more and more to form (statistical and
temporary) clusters the smaller they are. Since these clusters may
behave like single bigger particles \cite{sam00, sam01},
$\Gamma(W, R)$ should be especially for small $R$ smaller than the
fluidization acceleration expected from Eq. \ref{gamma-wr-2:eq}.
This is consistent with $\beta < 1$, as observed.

\begin{figure}[h]
 \includegraphics[width = 8.5cm]{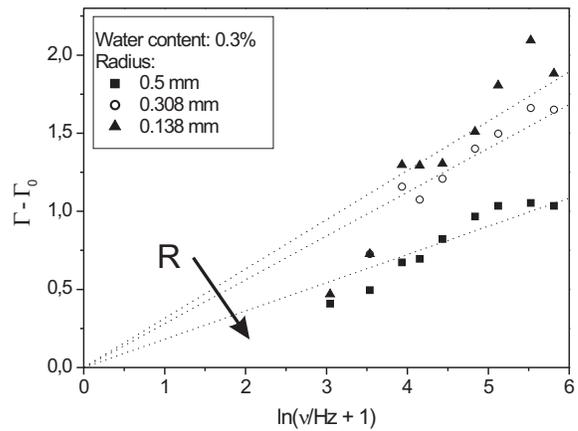}
 \caption{\label{gamma-nu:pict} Frequency dependence of the fluidization acceleration for
 different bead radii for $W = 0.3 \%$ where $\Gamma(W)$ has already
 saturated.
 The fitting functions are described in the text.}
\end{figure}

{\it Frequency dependence:} The fluidization amplitude turned out
to depend also on the shaking frequency $\nu$. In general we
observed that $\Gamma$ was a concave and monotonous rising
function of $\nu$. This frequency dependence was most distinctive
for small beads and at high liquid contents, although it did not
depend very strongly on $W$ for the bigger $R$. Some sample curves
of this dependence are shown in Fig. \ref{gamma-nu:pict}.

The frequency dependence could be fitted with the function
\[
\Gamma (\nu) = \Gamma_0 \left( 1 + \zeta \ln (1 + \nu/\nu_0)
\right)
\]
with
\[
\zeta := \frac{C_1 \widetilde{c} f(w)}{R^{\beta} R_0^{1 - \beta}}
\left( \frac{1}{H} + \frac{1}{H_c} \right)
\]
(see Eq. \ref{factorize:eq}) for the applied large frequencies
with the fitting parameter $\zeta$ (and accordingly $C_1$), as is
shown in Fig. \ref{gamma-nu:pict}.  The value of $\nu_0$ turned
out to be of the order of $1 Hz$. The frequency dependence was
more pronounced for the small beads, i.e. $\zeta$ was a monotonous
falling function of $\nu$. It emerged that $\zeta (R)$ was roughly
proportional to $R^{- \beta}$ (see Fig. \ref{gamma-r:pict}).

We will now provide a qualitative explanation for this behavior.
The amplitudes $A$ corresponding to the frequencies $\nu$ were
only about $1 \mu m$ for the higher $\nu$ which is much smaller
than the radii of the glass beads. This means that a translational
motion of the glass beads is more and more difficult for higher
frequencies and occurs only through many-particle collisions. Thus
the rotational degree of freedom becomes more and more important
for the higher frequencies which leads to a concave form of
$\Gamma (\nu)$, since rotational motion is not impeded
significantly by small $A$ if the particles are perfect spheres
\footnote{One important parameter for this dependence is the bead
roughness.}. We have already mentioned the occurrence of particle
clusters in the wet granulate. Since these clusters are irregular,
they cannot easily rotate. Hence one expects a more pronounced
frequency dependence, i.e. a bigger $\zeta$, for the smaller beads
where the assumed clusters are more pronounced (see above). This
is in qualitative agreement with our measurements.

The rotational motion of non-spherical particles in a granulate
consisting of them is much more difficult (as for the case of
clusters described above) and depends clearly on $A$. Hence we
conducted also some experiments with cubes \footnote{We used NaCl
with a size of $300 \mu m$; as wetting liquid we used nonane. (M.
Scheel, D. Geromichalos, S. Herminghaus; manuscript in
preparation)} and other particles like cylinders, river sand and
foraminiferes in order to check our assumption. For the case of
cubic particles \footnote{$W = 0.3 \%$; $50 Hz \leq \nu \leq 333
Hz$} we observed indeed a very different behavior: The increase of
$\Gamma (\nu)$ was nearly linear and quite strong \footnote{We
observed a linear increase from $\Gamma \approx 2$ at $50 Hz$ to
$\Gamma \approx 8$ at $333Hz$.}. The $\nu$-dependence of $\Gamma$
was for the other non-spherical particles also linear but not as
pronounced as for the cubical case. The dependence $\zeta_1 :=
\frac{\mbox{d}\Gamma(\nu)}{\mbox{d}\nu}$ in the case of cylinders
was very close to that for the glass beads, while the cases of
river sand and foraminiferes lied in-between the observed extremes
given by the spheres and the cubes. Concluding it can be said that
$\zeta_1$ increased with the 'roughness' (or 'edginess') of the
particles. Furthermore $\zeta_1$ was a monotonous rising function
of $W$ which is also consistent with our explanation of the
$\nu$-dependence \footnote{The liquid leads to a formation of
clusters. This causes an additional obstruction of rotational
motions and consequently an increase of the fluidization
acceleration.}.

It is interesting to note that for the beads, with a roughness
exponent $\chi \approx 0$, we get $\Gamma - \Gamma_0 \propto \ln
(\nu)$, while for edgy grains, with $\chi \approx 1$ we find
$\Gamma - \Gamma_0 \propto \nu^1$. Whether this finding represents
a coincidence or bears some fundamental physics remains to be
investigated.


Inspiring discussions with M. Schulz are gratefully acknowledged.
The authors thank the German Science Foundation for financial
support within the Priority Program `Wetting and Structure
Formation at Interfaces'.

\bibliography{apssamp}

\end{document}